\documentclass[12pt,preprint]{aastex}

\usepackage{color}
\definecolor{Blue}{rgb}{0.3,0.3,0.9}

\newcommand{\TEV}{TeV~J2032+4130}
\newcommand{\gr}{$\gamma$-ray}
\newcommand{\grs}{$\gamma$ rays}
\newcommand{\ch}{Cherenkov}

\newcommand{\myemail}{akonopel@purdue.edu}

\shorttitle{Unidentified Cygnus TeV Gamma-ray Source}
\shortauthors{VERITAS collaboration}

\begin{document}


\title{Observations of the unidentified TeV $\gamma$-Ray Source {\TEV} 
with the Whipple Observatory 10 m Telescope}

\author{
A.~Konopelko\altaffilmark{1},
R.W.~Atkins\altaffilmark{2},
G.~Blaylock\altaffilmark{3},
J.H.~Buckley\altaffilmark{4},
Y.~Butt\altaffilmark{5},
D.A.~Carter-Lewis\altaffilmark{6},
O.~Celik\altaffilmark{7},
P. Cogan\altaffilmark{8},
Y.C.K.~Chow\altaffilmark{7},
W.~Cui\altaffilmark{1},
C.~Dowdall\altaffilmark{8},
T.~Ergin\altaffilmark{3},
A.D.~Falcone\altaffilmark{9},
D.J.~Fegan\altaffilmark{8},
S.J.~Fegan\altaffilmark{7},
J.P.~Finley\altaffilmark{1},
P.~Fortin\altaffilmark{10},
G.H.~Gillanders\altaffilmark{11},
K.J.~Gutierrez\altaffilmark{4},
J.~Hall\altaffilmark{2},
D.~Hanna\altaffilmark{12},
D.~Horan\altaffilmark{13},
S.B.~Hughes\altaffilmark{4},
T.B.~Humensky\altaffilmark{14},
A.~Imran\altaffilmark{6},
I.~Jung\altaffilmark{4},
P.~Kaaret\altaffilmark{15},
G.E.~Kenny\altaffilmark{11},
M.~Kertzman\altaffilmark{16},
D.B.~Kieda\altaffilmark{2},
J.~Kildea\altaffilmark{12},
J.~Knapp\altaffilmark{17},
K.~Kosack\altaffilmark{4,*},
H.~Krawczynski\altaffilmark{4},
F.~Krennrich\altaffilmark{6},
M.J.~Lang\altaffilmark{11},
S.~LeBohec\altaffilmark{2},
P.~Moriarty\altaffilmark{18},
R.~Mukherjee\altaffilmark{10},
T.~Nagai\altaffilmark{6},
R.A.~Ong\altaffilmark{7},  
J.S.~Perkins\altaffilmark{13},
M.~Pohl\altaffilmark{6},
K.~Ragan\altaffilmark{12},
P.T.~Reynolds\altaffilmark{19},
H.J.~Rose\altaffilmark{17},
G.H.~Sembroski\altaffilmark{1},
M.~Schr\"{o}dter\altaffilmark{6},
A.W.~Smith\altaffilmark{13},
D.~Steele\altaffilmark{20},
A.~Syson\altaffilmark{17},
S.P~Swordy\altaffilmark{14},
J.A.~Toner\altaffilmark{11},
L.~Valcarcel\altaffilmark{12},
V.V.~Vassiliev\altaffilmark{7},
R.G.~Wagner\altaffilmark{21},
S.P.~Wakely\altaffilmark{14},
T.C.~Weekes\altaffilmark{13},
R.J.~White\altaffilmark{17},
D.A.~Williams\altaffilmark{22},
B.~Zitzer\altaffilmark{1}\
(The~VERITAS~Collaboration)
}

\email{<\myemail>}


\altaffiltext{1}{Department of Physics, Purdue University, 
West Lafayette, IN 47907, USA}

\altaffiltext{2}{Physics Department, University of Utah, 
Salt Lake City, UT 84112, USA}

\altaffiltext{3}{Department of Physics, University of Massachussetts,
Amherst, MA 01003-4525, USA}

\altaffiltext{4}{Department of Physics, Washington University 
in St. Louis, St. Louis, MO 63130, USA}

\altaffiltext{5}{SAO, 60 Garden St., Cambridge, MA 02138}

\altaffiltext{6}{Department of Physics and Astronomy, Iowa State
University, Ames, IA 50011, USA}

\altaffiltext{7}{Department of Physics and Astronomy, University of
California, Los Angeles, CA 90095, USA}

\altaffiltext{8}{School of Physics, University College Dublin,
Belfield, Dublin 4, Ireland}

\altaffiltext{9}{Department of Astronomy and Astrophysics, Penn State
University, University Park, PA 16802, USA}

\altaffiltext{10}{Department of Physics and Astronomy, 
Barnard College, Columbia University, NY 10027, USA}

\altaffiltext{11}{Physics Department, National University of Ireland,
Galway, Ireland}

\altaffiltext{12}{Physics Department, McGill University, Montreal, QC
H3A 2T8, Canada}

\altaffiltext{13}{Fred Lawrence Whipple Observatory,
Harvard-Smithsonian Center for Astrophysics, Amado, AZ 85645, USA}

\altaffiltext{14}{Enrico Fermi Institute, University of Chicago,
Chicago, IL 60637, USA}

\altaffiltext{15}{
Department of Physics and Astronomy, University of Iowa, 
Iowa City, IA 52242 USA}

\altaffiltext{16}{Department of Physics and Astronomy, DePauw
University, Greencastle, IN 46135-0037, USA}

\altaffiltext{17}{School of Physics and Astronomy, University of Leeds,
Leeds, LS2 9JT, UK}

\altaffiltext{18}{Department of Physical and Life Sciences,
Galway-Mayo Institute of Technology, Dublin Road, Galway, Ireland}

\altaffiltext{19}{
Department of Applied Physics and Instrumentation, 
Cork Institute of Technology, Bishopstown, Cork, Ireland}

\altaffiltext{20}{Astronomy Department, Adler Planetarium and 
Astronomy Museum, Chicago, IL 60605, USA}

\altaffiltext{21}{
Argonne National Laboratory, 9700 S. Cass Avenue, 
Argonne IL 60439, USA}

\altaffiltext{22}{Santa Cruz Institute for Particle Physics and
Department of Physics, University of California, Santa Cruz, CA 95064,
USA}

\altaffiltext{*}{currently at the Max-Planck-Institute of Nuclear Physics, 
Saupfercheckweg 1, 69117 Hedidelberg, Germany}




\begin{abstract}
We report on observations of the sky region around the unidentified 
TeV {\gr} source (TeV~J2032+4130) carried out with the Whipple Observatory 
10~m atmospheric Cherenkov telescope for a total of 65.5~hrs between 2003 
and 2005. {{The standard two-dimensional analysis developed by the Whipple 
collaboration for a stand-alone telescope reveals 
an excess in the field of view at a {pre-trials} significance 
level of 6.1$\sigma$. 
The measured position of 
this excess is
$\alpha_{2000}=20^h 32^m 27^s$, $\delta_{2000}= 41^\circ 39' 17''$. 
The estimated integral flux for this {\gr} source  
is about 8\% of the Crab-Nebula flux. 
{The data are consistent with a point-like source.}
Here we present a detailed description 
of the standard two-dimensional analysis technique used for the 
analysis of data taken with the Whipple Observatory 10~m telescope 
and the results for the {\TEV} campaign. We include a short 
discussion of the physical mechanisms that may be responsible for 
the observed {\gr} emission, based on possible association 
with known astrophysical objects, in particular Cygnus~OB2.}}

\end{abstract}





\section{Introduction}

During observations of the Cygnus~X-3 region in 1993 by the Crimean
Astrophysical Observatory, using the GT-48 imaging atmospheric
{\ch} telescope, a serendipitous source at a pre-trial 
significance of 6$\sigma$ was detected at a position of approximately 
0.7$^\circ$ to the north of Cygnus~X-3. Assuming an integral spectral 
index of -1.5, \citet{nes95} reported the {\gr} flux of
this unidentified source above 1~TeV as $3\times \rm 10^{-11}~cm^{-2}s^{-1}$,  
which is about 1.7 times the Crab-Nebula flux.

Independent observations of the Cygnus~X-3 region with the High
Energy Gamma Ray Astronomy (HEGRA) system of five imaging atmospheric
{\ch} telescopes deployed
at La~Palma, Canary Islands, were performed during 1999-2001 with
10 milliCrab sensitivity and arc-minute resolution. These
observations revealed a region of extended {\gr} emission at a
significance level of $\sim$5$\sigma$ \citep{aha02} that is
positionally consistent with the {\gr} source originally detected
by the Crimean
Astrophysical Observatory. Follow-up observations of this unidentified TeV
{\gr} source in the Cygnus region with HEGRA in 2002 enabled
a rather accurate measurement of the source position,
$\alpha_{2000}=20^h 31^m 57^s$ $\delta_{2000}= 41^\circ 29'
56.8''$, and its angular extent, $6.2'\pm 1.2'_{stat}\pm
0.9'_{sys}$ \citep{aha05a}. The {\gr} flux above 1~TeV reported by
\citet{aha05a} was $\rm (6.89\pm
1.83)\times10^{-13}~cm^{-2}s^{-1}$, which is $\sim$5\% of the Crab
Nebula flux. The source has a power-law energy spectrum with a
hard photon index of $-1.9\pm0.1_{stat}\pm0.3_{sys}$.

Cygnus X-3 was the focus of extensive observations with the Whipple Observatory 10~m imaging
atmospheric {\ch} telescope during 1989-1990. 
There was no evidence of a signal from Cygnus X-3 \citep{ofla}.
A total of
50.4~hrs of analyzable data were accumulated during that campaign. 
These observations included 
in the field of view the
reported location of {\TEV}. An analysis of these archival data by
\citet{lang04} resolved an excess of emission close to the HEGRA position of
{\TEV} at a significance level of 3.3$\sigma$. It is worth noting
that the peak signal in the Whipple Observatory data was
noticeably offset by $\sim 3.6'$ to the north-west of the
HEGRA source position. \citet{lang04} reported the {\gr} flux of
{\TEV} to be 12\% of the Crab-Nebula flux above 400 GeV.

There are presently no well-established counterparts of {\TEV} at
other wavelengths \citep{butt06} despite the fact that the source is located within
the bounds of the Cygnus OB2 association \citep{aha05a}, an active 
star-forming region. As such, {\TEV} represents a new class of the TeV {\gr}
sources commonly referred to as {\it dark accelerators} owing to their unknown origin.

\section{Experiment}


The Whipple 10~m atmospheric Cherenkov telescope
consists of a cluster of photomultiplier
tubes placed at the focus of a relatively large optical reflector. The images
of the {\ch} light flashes generated both by {\gr} and charged cosmic-ray
primaries interacting in the Earth's atmosphere are digitized and recorded. A 
dedicated off-line analysis of these
images enables a substantial suppression of the large cosmic-ray background
and therefore dramatically improves the resulting signal-to-noise ratio.

The reflector of the Whipple Observatory imaging atmospheric
{\ch} telescope is a tessellated
structure consisting of 248 spherical mirrors, which are hexagonal in shape 
and 61~cm size from apex to apex, arranged
in a hexagonal pattern \citep{caw90}. The mirrors are mounted on a steel
support structure, which has a 7.3~m radius of curvature 
with a 10~m aperture. Each
individual mirror has $\sim$14.6~m radius of curvature and is pointed toward
a position along the optical axis at 14.6~m from the reflector.
This arrangement constitutes a \citet{dc57} design of the optical reflector.
The point-spread function of the Whipple Observatory 10~m telescope has a
FWHM of $\sim 7.2'$ on-axis.

In 1999, a 490-pixel high-resolution camera (GRANITE III) was installed at the
Whipple Observatory \citep{jf01}. It consists of an inner camera of 379 PMTs 
in a close-packed hexagonal arrangement (each PMT subtending $0.11^\circ$ on 
the sky) and has a $2.6^\circ$ diameter. The inner camera is surrounded by 
111 PMTs of $0.24^\circ$ in 3 concentric rings.
The overall field of view of the camera is $4.0^\circ$ in diameter.
{However, the 3 concentric rings of $0.24^\circ$ pixels 
were removed from the camera in 2003 so that only the 379 inner pixels 
were present during the {\TEV} campaign.}  
A set of
light concentrators is mounted in front of the inner pixels to increase the 
light-collection efficiency by $\sim38$\%. The camera triggers if the signal in each
of at least 3 neighboring PMTs out of the inner 331 exceeds a threshold of 32~mV,
corresponding to $\sim$8-10~photoelectrons. The post-GRANITE III 
upgrade trigger rate of the Whipple Observatory 10~m telescope
is $\sim 20-30$~Hz at zenith. The recorded images are first flat-fielded
using nightly measured nitrogen arc lamp PMT responses and then cleaned
by applying a standard {\it picture} and {\it boundary} technique with thresholds of
4.25 and 2.25 times the standard deviation of the PMT pedestal distributions,
respectively (see, e.g., \citealt{jk06}). To characterize the shape and
orientation of calibrated images, the standard second-moment parameters
are calculated as described by \citet{rey93}. To equalize the night-sky noise
in the {\it ON} and {\it OFF} fields, a software padding technique (see, e.g.,
\citealt{rl01}) is applied.

The response of the Whipple 10~m telescope has changed over time due primarily 
to degradation of the optical elements, occasional readjustment of the PMT 
gains and seasonal
variations of atmospheric transparency. Fortunately the telescope response
(e.g., event-detection rate, distribution of image sizes) to the steady
cosmic-ray flux is extremely sensitive to each of these effects and it can
be effectively used for validating the actual telescope performance.
{\citet{leb03} developed a standard procedure to use cosmic-ray events 
taken at the zenith to track changes in the  
instrument {\it throughput} 
that reflect changes in the instrument sensitivity over time.}
This {\it throughput} factor can be
measured using the luminosity distribution of the recorded cosmic-ray
flashes and it allows accurate monitoring of the telescope response
throughout periods of observation not affected by major hardware upgrades.
A somewhat similar approach was used earlier with the first stand-alone HEGRA
telescope \citep{ak96}.

\section{Observations}

The position of {\TEV} was observed with the Whipple Observatory 10~m
imaging atmospheric {\ch} telescope at Mt. Hopkins for about 65~hrs of 
good on-source data between 2003 and 2005. 
{ Data were obtained in the {\it ON/OFF} mode where each 
{\it ON}-source data run is either immediately preceded or
followed by a matching {\it OFF}-source run where the telescope 
tracks the same region of zenith angles but with an offset in 
right ascension from the true source position.}
The observations were taken in pairs
of both ``{\it ON} before {\it OFF}'' and ``{\it OFF} before {\it ON}'' 
runs of 28~min duration each. This practice 
provided two independent background fields to help minimize 
systematic effects due to
the bright sky in the vicinity of \TEV. To further reduce any possible
systematic bias in the on-source sample of recorded images, caused by
inhomogeneous illumination across the camera field of view, a 
fraction of observational data was taken using 38~min ``{\it ON} before 
{\it OFF}'' 
and ``{\it OFF} before {\it ON}'' runs. Thus the total data set employed 4 
independent background fields
to minimize sky-brightness systematics.
{\TEV} was observed during four nearby epochs between 2003 and 2005
(see Table~\ref{data_summary}). A total of 132 data pairs were
collected in good weather at zenith
angles less than $30^\circ$. The average elevation and the average 
{\it throughput}
factors for the four observational periods are given in
Table~\ref{data_summary}.

\begin{figure}
\epsscale{0.7}
\plotone{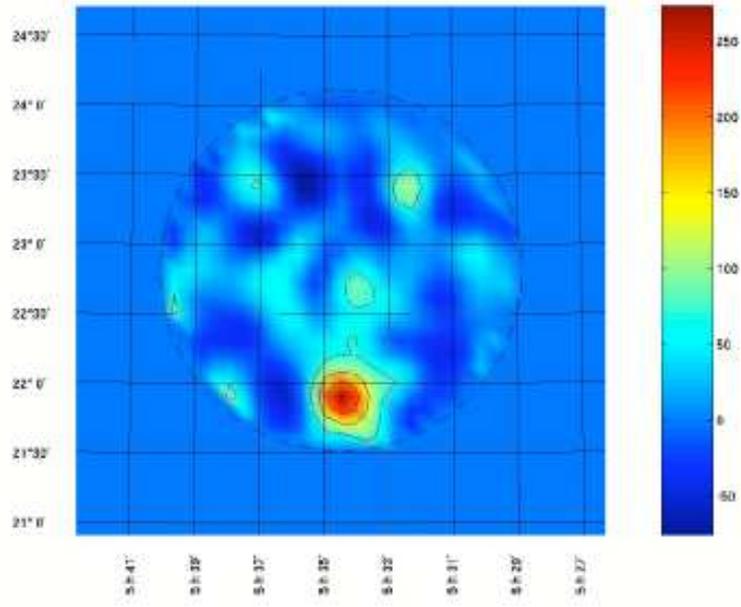}
\caption{{{Gaussian-smoothed ($\sigma=7.6'$) declination vs right ascension 
map of the excess counts from the Crab Nebula observed 
with the 0.8$^\circ$ offset towards north from the telescope optical axis  
for a total of 3.2~hrs. The color bar 
represents the excess counts and the coordinates are referenced to the epoch J2000.}} 
\label{crab08off}}
\end{figure}

The 10~m telescope has custom tracking software that has been in use for
approximately 10 years.  {The 10~m telescope pointing model has been determined by 
imaging stars on a white screen mounted at the focal plane to measure 
pointing errors as a function 
of azimuth and elevation. These pointing errors are used to develop 
the corresponding T-point corrections. Typically the 
T-point corrections are done at intervals of about three months, 
with an error between subsequent corrections typically less than $6'$.}
T-point corrections are applied to the tracking software 
to account for gravitational flexure of the structure as a function 
of the azimuth and elevation of the telescope. To monitor the tracking of 
the telescope during routine data taking, {\it tracking records} are stored once 
every 30 seconds in the data stream. These records include the position 
of the pointing 
direction at the current epoch, the canonical position of the source at the 
current epoch, and the azimuth and elevation of the telescope 
derived from the telescope encoders. 
These data allow us to check that the pointing direction of the telescope are 
consistent from run to run and season to season as we accumulate a database of 
long observations on a particular source. 
{{We have examined the 
results of comparing the encoder-derived azimuth and elevation of the source 
under study here, \TEV, and the pointing direction of the telescope. The 
pointing direction is consistent with the source direction from season to 
season and any offset is much smaller than the size of the central PMT.}} 
Additionally, pointing checks are acquired on a routine basis by placing 
a bright star (in the vicinity of the source under study) at the center 
of the field of view and recording the PMT currents. These pointing
checks indicate an absolute offset of $~3^{\prime}$ (i.e., less than half
the single-pixel field of view) and are consistent with the offline analysis 
of Crab-Nebula data {{(see Figure~\ref{crab08off}).}} 

\section{Data Analysis}
\label{2D}
The data-analysis pipeline consists of two distinct phases. The data are
first processed and distributions from the raw uncut data are accumulated as 
diagnostics of both the condition of the instrument and the stability of the
weather conditions. Each data run is visually inspected for rate stability, 
timing
stability and tracking consistency, and either accepted or rejected based on this
first pass. Once this diagnostic pass is made, 
acceptable runs are further processed
for scientific investigation.
Despite the significant advancements that have been made in different
aspects of the imaging atmospheric {\ch} technique during the last
decade, a canonical analysis method known as {\it Supercuts} \citep{pu91}
still stands as the most effective set of {\gr} image-selection criteria
for the Whipple Observatory 10~m telescope. This method utilizes both
the shape and orientation information in the recorded {\ch} light images
\citep{fe97}. The choice of optimal analysis cuts heavily relies on the
actual configuration of the imaging camera, e.g. the angular size of PMTs,
total field of view, the level of night-sky background light in each
pixel etc. Therefore after the recent hardware upgrade for GRANITE~III
had been completed, a new set of {\it Supercuts} was developed in 2001
using a Crab-Nebula data sample that was rich in $\gamma$-ray content 
(see Table~\ref{supercuts}). Since
then, this particular implementation of {\it Supercuts} has remained the
standard selection method for subsequent data taken with the Whipple
Observatory 10~m telescope.

In an {\it a priori} search for point-like {\gr} sources, the standard
{\it Supercuts} includes an orientation parameter, $\alpha$, in addition
to the parameters listed in Table~\ref{supercuts}. In the present 
investigation, we used instead
a two-dimensional analysis described previously by \citet{b98} and 
\citet{rl01} for off-axis
or extended {\gr} sources. Images of the {\gr} showers have
their major axes preferentially pointed towards the source position
on the sky. The elongation of an image, which commonly has an elliptic
shape, defines a point of origin for that individual event.
For a source of {\grs} positioned anywhere within the camera
field of view, the shower images will point towards that actual source
position in the camera. The angular distance from the image
centroid (the center of gravity) to the point of origin can be
determined as
\begin{equation}
d = \zeta (1-Width/Length) = \zeta \xi,      \label{lessard}
\end{equation}
where $Width$ and $Length$ are the transverse and lateral
angular extensions of the image, respectively \citep{fe97}. 
$\xi$ is an ellipticity
parameter of the image, which is by definition equal to 0 for a
circular image. Note that $\zeta$ is the only free parameter in
Eqn.(\ref{lessard}). The straight line along the major axis of the image
can be rendered in a Cartesian coordinate system on the camera focal plane
using the position of the image centroid
and the azimuthal angle of the image.
The angular distance
along this line from the image centroid to the point of origin can be
computed using Eqn.(\ref{lessard}), which ultimately determines a
unique arrival direction for every recorded shower. A large set of 
Crab-Nebula data was used to derive that optimal value of $\zeta$ parameter
which provides a minimal spread of the points of origin around the known
position of a point-like {\gr} source. Analysis of the Crab-Nebula data
yields an optimal
value of $\zeta =1.37$. The resulting precision
for directional shower reconstruction with this optimal $\zeta$ is 
$\sigma \simeq 7.8'$. 
Source localization for a bright {\gr} source (1 Crab) is of the order 
of a few arcmins after 1~hr observing time and is comparable to the systematic 
uncertainty in the telescope pointing, about $4'$.

In the two-dimensional analysis of images recorded by the Whipple
Observatory 10~m telescope, all calibrated, cleaned and
parameterized events in the {\it ON} and {\it OFF} data sets are analyzed,
first, with {\it Supercuts} (see Table~\ref{supercuts}) and 
consequently binned in a two-dimensional grid, mapping the sky field
around the position tracked by the telescope. There were three major
approaches used to perform a two-dimensional analysis.
In particular, one can generate (i) a sky map (declination vs right ascension) 
of uncorrelated rectangular
bins with an angular size of $0.1^\circ \times 0.1^\circ$ ; (ii)
a sky map smoothed with a circular aperture of $0.22^\circ$ radius,
and (iii) a Gaussian-smoothed sky map, in which each candidate
{\gr} event receives a statistical weight of
\begin{equation}
\omega = 1/\sqrt{2 \pi \sigma_o}e^{-((\theta_x - \theta^*_x)^2+
(\theta_y - \theta^*_y)^2)/\sqrt{2 \pi \sigma_o}}, \label{gaussian}
\end{equation}
where $\sigma_o$ is the actual width of the telescope point-spread function,
derived from Crab-Nebula observations, and $(\theta^*_x,\theta^*_y)$
is the current reference position within the grid. By subtracting the
number of counts in the {\it OFF} map from the corresponding number 
of counts in the {\it ON} map, one can compute the excess
in recorded events for each position within the camera field
of view covered by the grid. Excess counts in this difference map 
represent the 
number of {\grs} from the putative source. Due to truncated events 
(i.e., events that are not contained within the fiducial area of the camera) 
and the front/back ambiguity 
of the two-dimensional analysis \citep{rl01}  we restrict the 
field of view for the analysis to a radius of $1.25^{\circ}$ from the 
telescope pointing direction. This restriction minimizes systematics 
resulting from events with their light distribution close to the edge and 
external to the camera field of view. 
Note that the actual observing time in the {\it ON} and {\it OFF} 
fields might not be
identical: this is taken into account by applying the procedure 
described by \citet{lima83}. 

\section{Telescope Performance}

\begin{figure}
\epsscale{0.95}
\plotone{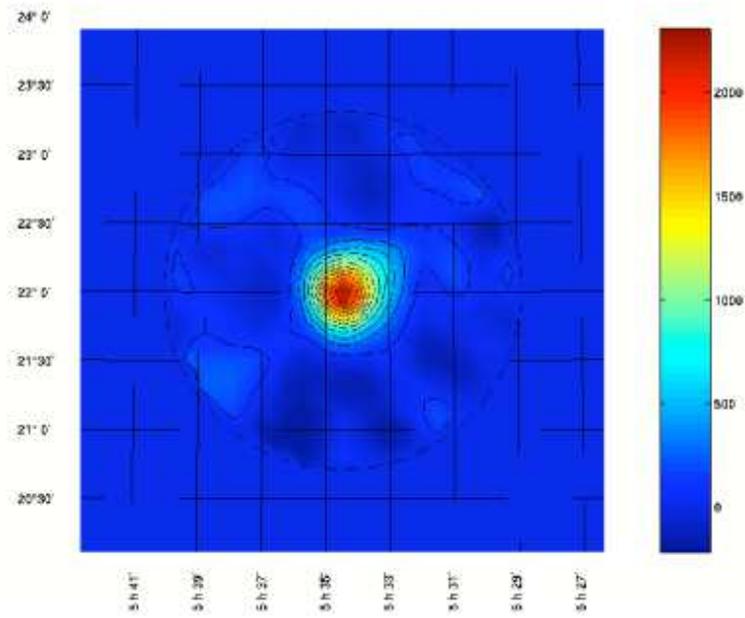}
\caption{Gaussian-smoothed ($\sigma=7.6'$) declination vs right ascension 
map of the excess counts from the Crab-Nebula region observed
for a total of 18.7~hrs in 2005. The color bar represents the excess counts and the coordinates are referenced to the epoch J2000. 
\label{fg2}}
\end{figure}

The performance of the Whipple Observatory 10~m telescope during
2003-2005 can be estimated using contemporaneous observations of the
Crab Nebula, which is the standard candle of VHE {\gr} astronomy
\citep{tcw89}. The Crab Nebula was routinely observed
with the Whipple Observatory 10~m telescope for normalization of
the instrumental response during three consecutive epochs during the
winter seasons of
2003/2004, 2004/2005 and 2005/2006, for 13.8~hr, 12.2~hr and 18.7~hr,
respectively. The complete data-reduction chain described here was
tested in great detail on the Crab-Nebula data. The two-dimensional
sky maps of extracted candidate {\gr} events were generated separately
for the {\it ON} and {\it OFF}
data sets
for a similar
$1.25^\circ \times 1.25^\circ$ field of view with
$0.1^\circ \times 0.1^\circ$  uncorrelated rectangular bins.
These two-dimensional sky maps have been used to produce the excess-counts
map. An excess of very high statistical
significance is seen at the position of the Crab Nebula. The
angular shape of the {\gr} signal from the Crab Nebula can be well
reproduced by the two-dimensional Gaussian
\begin{equation}
f(\theta_x,\theta_y) =
A_o e^{-\frac{1}{2}(\theta_x- \bar{\theta}_x)^2/\sigma_x^2}
e^{-\frac{1}{2}(\theta_y-\bar{\theta}_y)^2/\sigma_y^2}, \label{gauss}
\end{equation}
where two sets of parameters, ($\bar{\theta}_x$,$\bar{\theta}_y$) and
($\sigma_x$,$\sigma_y$), characterize the systematic offset and broadness
of the point-spread function, respectively.
The parameters of the fit obtained for three
nearby observational epochs are summarized in Table~\ref{fit}. Note
that the position of the {\gr} peak deviates from the position of the
Crab Nebula by less than $3'$. The average width of the point-spread 
function is $\sigma \simeq 7.6'$ (see Figure~\ref{fg2}). 
This observationally determined $\sigma$
is used as the width of the
Gaussian distribution invoked in Eqn.(\ref{gaussian}), which was
adopted for the smoothing of the two-dimensional sky maps. For an additional
crosscheck, a number of {\it ON} and {\it OFF} Crab-Nebula pairs were taken
with a $0.5^\circ$ and $0.8^\circ$ offsets from the nominal position 
{{(see Figure~\ref{crab08off})}}. 
These data runs were analyzed using exactly the same
two-dimensional analysis method as described above and the resulting sky maps of the
Crab-Nebula region show a clear {\gr} excess displaced
from the center of the field of view. The position of the Crab-Nebula {\gr}
peak is found to be consistent with the initial offset 
and the width of the {\gr} signal distribution is the 
same size as for the {\it ON}-axis observations.

To determine the position and angular extent of a 
putative {\gr} source in the field of view of the 10~m telescope 
the two-dimensional excess counts map is fitted to a model of a Gaussian 
{\gr} brightness profile of the form 
$p \propto exp(-\theta^2/2(\sigma^2_{source}+\sigma^2_{PSF}))$, where 
$\sigma_{source}$ and $\sigma_{PSF}$ are the approximate angular size 
of the source and the width of the point-spread function, respectively. 
The origin of the Gaussian fit determines the source position.

The two-dimensional analysis of the Crab-Nebula data taken during three
consecutive observing periods yields a rather stable {\gr} rate and
signal-to-noise ratio (see Table~\ref{crab_data}). { Some 
remaining seasonal variations can be attributed to changes of the 
telescope response corresponding to various hardware conditions such as gain
change of the PMTs, mirror reflectivity, etc.} 
 After applying
{\it Supercuts} and an aperture cut of $0.22^\circ$, the measured Crab
{\gr} rate was $\sim$1.0~$\gamma$/min, which corresponds to a
signal-to-noise ratio of about 4~$\sigma/\sqrt{hr}$
(see Table~\ref{crab_data}). It is worth noting that the one-dimensional
analysis utilizing the $\alpha$ parameter yields a higher signal-to-noise
ratio as well as a correspondingly higher {\gr} rate. {This is due 
to a
front/back ambiguity of the arrival direction determination. Normally this 
ambiguity is resolved by measuring the asymmetry of the light distribution in 
the image to choose the ``correct'' arrival direction. However, the small 
field of view of the camera utilized for this data sample, $2.4^{\circ}$ in 
diameter, prevents us from making a reasonable estimate of the asymmetry. We 
are then forced to accept both solutions for the arrival direction, front and 
back, for a given image orientation in the focal plane.}

The sensitivity of the Whipple Observatory 10~m telescope to a {\gr}
signal within its field of view  can be noticeably improved by applying
Gaussian smoothing to the {\it ON} and {\it OFF} sky maps. In this approach
(see Section~\ref{2D}) each of the events accepted by {\it Supercuts}
obtains a statistical weight that is assessed according to the events
angular distance from the current test position for a {\gr} source. A
two-dimensional Gaussian distribution, 
centered at the test source position and with width
$\sigma=7.6'$ along each dimension of the Cartesian coordinate
system, determines the statistical
weight of the candidate {\gr} events. The performance of this method was
evaluated with the Crab-Nebula data sample (see Figure~\ref{fg2}) and the 
summary of these results is given in Table~\ref{crab_data}. This method
yields a substantial recovery in the {\gr} rate and a correspondingly
higher significance of the excess. An analysis of the Gaussian-smoothed
two-dimensional maps yields results which are comparable with those
derived from a standard $\alpha$ analysis of Crab-Nebula data taken in
the {\it ON}-source observation mode (see Table~\ref{crab_data}).

\section{Results}

\begin{figure}
\epsscale{0.7}
\plotone{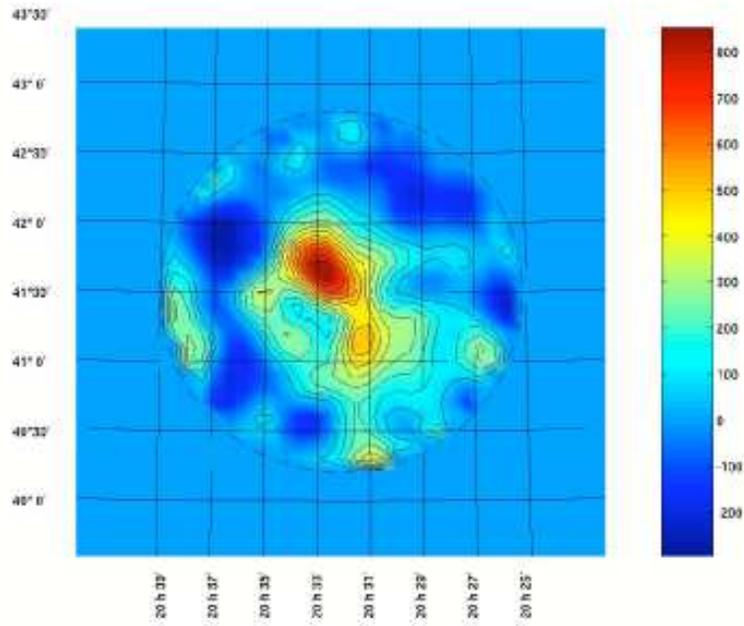}
\caption{Gaussian-smoothed ($\sigma=7.6'$) declination vs right ascension 
map of the excess counts from the {\TEV} region observed for a
total of 65.5~hrs between 2003-2005. The contours are in steps of
0.5~$\sigma$ of the signal significance.
\label{fg3}}
\end{figure}

The {\TEV} observational data taken with the Whipple Observatory
10~m telescope at Mt.~Hopkins between 2003 and 2005 in the 
{\it ON}/{\it OFF} mode
for a total {\it ON}-source time of 65.6~hrs have been analyzed using the
standard analysis methods developed by the Whipple collaboration. These
methods have been tested in great detail for various well-established
{\gr} sources detected with the Whipple Observatory 10~m telescope,
particularly the Crab Nebula, which is a standard candle of ground-based
TeV {\gr} astronomy. The two-dimensional Gaussian-smoothed excess-counts
map of the {\TEV} sky region shows {{a distinct excess}} 
in the vicinity
of the HEGRA unidentified {\gr} source (see Figure~\ref{fg3}). {
{The significance of this excess and its celestial coordinates
are summarized in Table~\ref{tev2032}.}} {{This excess seen by the
Whipple Observatory 10~m telescope (see Figure~\ref{fg3}) has
an angular displacement of about $9'$ from the HEGRA {\gr}
source. Given the statistical and systematic uncertainties in the source 
localization (about $4'$ and $6'$, respectively), the displacement of 
the excess in Figure~\ref{fg3} is consistent 
with the position of the HEGRA unidentified {\gr} source.
{ A Gaussian fit of the smoothed excess-counts map shown 
in Figure~\ref{fg3} gives $\sigma_s=12.8'$ as the width of the excess. 
For comparison the correponding width of the $\gamma$-ray excess from 
the Crab Nebula is $\sigma_{Crab} = 11.4'$.}  
{ Based on these data, there is good statistical evidence 
for a source near the HEGRA detection and with angular extent less than $6'$.}
{A detailed analysis of the map shown in Figure~\ref{fg3} 
reveals the presence of a second excess located to the south-west of the 
HEGRA unidentified {\gr} source. However, the statistical significance of 
this excess corrected for the number of trials  
does not reach the 3$\sigma$ level which precludes our determination
of the nature of this enhancement as a $\gamma$-ray source.
}
Follow-up observations of the {\TEV} field with the 
VERITAS system of four 12~m 
atmospheric {\ch} telescopes { with substantially improved 
angular resolution} will allow us to carry out a dedicated search for 
possible extended {\gr} emission in the {\TEV} surroundings at a significantly 
improved sensitivity level.}}



{{
Based on the data reported here the source seen with 
 the Whipple Observatory 10~m telescope
is consistent with a point-like
{\gr} source. At the same time, given a $\sigma = 7.6'$ width of the PSF for the 10~m 
Whipple 
collaboration telescope, we can not distinguish between a point source and 
a diffuse source with extent less than a $6'$. Thus the Whipple source is 
consistent with HEGRA source in terms of its extension.}}



{{
The present Whipple Observatory signal for the {\gr} source
resolved in the vicinity of the HEGRA unidentified {\gr} source does
not have sufficient strength for 
adequate 
measurement of its
{\gr} spectrum. Assuming that the spectral shape of the emission is similar
to the standard candle {\gr} source, the Crab Nebula, one can 
estimate its {\gr} flux 
based on derived {\gr}
rates. Based on that assumption, the {\gr} flux is at the level of $\sim$8\%
of the Crab Nebula.}}
Assuming the source is at a distance D = 1.7~kpc, which is the 
distance to the Cygnus OB2 complex, its luminosity in TeV $\gamma$ rays is 
$$\rm L_{\gamma} \simeq 4\times 10^{33} (D/1.7\, kpc)^2 
(Flux/0.08\, Crab)\,\,\, erg\,s^{-1}.$$  
Although the {\gr} fluxes measured by different
groups suggest a steady {\gr} emission, a variable or sporadic nature of the
{\gr} emission from this source can not be ruled out at this stage given
the large uncertainty in the flux estimates.

\section{Discussion}

The {\TEV} HEGRA source belongs to a class of {\gr} sources known as
{\it dark accelerators}. These are presumably galactic sources owing to
their low galactic latitudes
and the lack of variability in TeV {\grs}. They have no compelling counterparts
at other wavelengths. Recently the High Energy Stereoscopic System (H.E.S.S.)
collaboration has discovered a population of unidentified {\gr} sources in
the Galactic plane \citep{aha05b, aha05c, aha06}. The underlying nature
of these sources is presently poorly understood. For instance, HESS~J1303-631,
which is the brightest among the unidentified {\gr} sources, could be
plausibly interpreted as the remnant of a {\gr} burst that occurred in our
Galaxy a few tens of thousands of years ago \citep{abk06}.

The TeV {\gr} emission observed by Crimean
Astrophysical Observatory \citep{nes95}, HEGRA \citep{aha05a}
and the Whipple Observatory \citep{lang04}, and the {\gr} emission
reported here, are located within the bounds of the Cygnus OB2 stellar
association \citep{aha02}. It is 1.7~kpc away, rather compact (about
$2^\circ$ across) and the most massive OB association known in the Galaxy,
implying a tremendous mechanical power density accumulated in the
stellar winds of its $\sim$2600 OB star members \citep{lpf02}. Such an
association offers a unique case to test the hypothesis of Galactic
cosmic-ray acceleration by the supersonic stellar winds of many young OB
stars propagating into the interstellar medium \citep{cp80,cm83}. In this
scenario, the TeV {\grs} can be the tracers of the
$\pi^o \rightarrow \gamma \gamma$ emission originating in the interactions
of very energetic nuclei with interstellar matter. Steady MeV-GeV
{\gr} emission detected by the EGRET instrument from the Cygnus OB2 region
(3EG~J2033+4118, \citealt{h99}) generally supports such a physical
interpretation.


Detection of the 
X-ray emission resolved from the {\gr} emitting region might help
to constrain severely the origin of the {\gr} emission, specifically helping
to determine whether electrons or nuclei are responsible for the production
of the TeV {\grs} seen from the Cygnus region. Recent observations 
of the unidentified TeV source in the Cygnus region  with the
{\it Chandra} satellite
revealed no obvious X-ray counterpart \citep{mu03,butt03,butt06}, 
evidently favoring a hadronic origin for the {\grs} from the Cygnus region. {
{However, it is
worth noting that 
the {\gr} emission region reported here 
(see Figure~\ref{fg3}) lies outside of the {\it Chandra} observational
window.}} It is apparent that further X-ray observations of a relatively broad
region around Cygnus 
could possibly provide a detection of the X-ray counterpart(s) and
consequently help to elucidate the physics of the 'dark accelerators' seen
in TeV {\grs}.

Future dedicated observations of the Cygnus region with advanced
ground-based (e.g. VERITAS) and satellite-borne (GLAST) {\gr} detectors
are required to help us understand the physics of this
population of unidentified galactic TeV {\gr} sources.

\section*{Acknowledgements}

\noindent
This research is supported by grants from the Smithsonian Institution, 
U.S. DOE, NSF, PPARC (UK), NSERC (Canada), and SFI (Ireland).

\begin{deluxetable}{clcccc}
\tablecaption{Summary of Data\label{data_summary}}
\tablewidth{0pt}
\tablehead{
\colhead{Epoch} & \colhead{Calendar period} & \colhead{{\it ON} time [min]} &
\colhead{Number of runs} & \colhead{Elevation} & \colhead{{\it Throughput}} }
\startdata
1 & Sep - Nov 2003 & 1471 & 54 & 73$^\circ$ & 1.01 \\
2 & Apr - Jun 2004         & 991  & 36 & 72$^\circ$ & 1.01 \\
3 & Sep - Nov 2004 & 525  & 15 & 75$^\circ$ & 1.08 \\
4 & May - Jul 2005           & 950  & 27 & 70$^\circ$ & 0.98 \\
\enddata


\end{deluxetable}

\begin{deluxetable}{lll}
\tablecaption{{\it Supercuts} selection criteria. \label{supercuts}}
\tablewidth{10cm}
\tablehead{\colhead{} & \colhead{Image parameter cut}}
\startdata
Trigger & Brightest pixel         $>$ 30~dc\tablenotemark{a} \\
        & Second brightest pixel  $>$ 30~dc\tablenotemark{a} \\
Shape   & $0.05^\circ < Width < 0.12^\circ$ \\
        & $0.13^\circ < Length < 0.25^\circ$ \\
Muon rejection & $Length/Size < 0.0004$ $\rm (^\circ/dc$\tablenotemark{a}) \\
Image quality  & $0.4^\circ < Distance <  1.0^\circ$
\enddata
\tablenotetext{a}{digital counts}

\end{deluxetable}

\begin{deluxetable}{lccccc}
\tablecaption{Parameters of the two-dimensional Gaussian fit
(Eqn.~\ref{gauss}) to the excess of {\gr} events from the
Crab Nebula observed during three observing epochs.
\label{fit}}
\tablewidth{0cm}
\tablehead{\colhead{Obs. period} & \colhead{$A_o$ (counts)} &
\colhead{$\bar{\theta}_x$ ($^\circ$)} & \colhead{$\bar{\theta}_y$ ($^\circ$)}
& \colhead{$\sigma_x$ ($^\circ$)} & \colhead{$\sigma_y$ ($^\circ$)} }
\startdata
2003/2004 & 124 & -0.028 & -0.026 & 0.120 & 0.131 \\
2004/2005 & 96  & -0.033 & -0.017 & 0.121 & 0.140 \\
2005/2006 & 154 & -0.037 &  0.001 & 0.138 & 0.106 \\
\enddata
\end{deluxetable}

\begin{deluxetable}{llcccccc}
\tablecaption{Summary of the Crab-Nebula data taken during three
observing epochs. S stands for the signal-to-noise ratio measured
in standard deviations of the excess cosmic-ray counts.
\label{crab_data}}
\tablewidth{0cm}
\tablehead{
\colhead{Obs. period} &
\colhead{t (min)} &
\colhead{ S\tablenotemark{a} ($\sigma$)} &
\colhead{$R_\gamma$\tablenotemark{a} ($\rm min^{-1}$)} &
\colhead{S\tablenotemark{b} ($\sigma$)} &
\colhead{$R_\gamma$\tablenotemark{b} ($\rm min^{-1}$)} &
\colhead{S\tablenotemark{c} ($\sigma$)} &
\colhead{$R_\gamma$\tablenotemark{c} ($\rm min^{-1}$)}
}
\startdata
2003/2004 & 828  & 21.1 & 3.01 & 16.0 & 1.18 & 22.3 & 2.6 \\
2004/2005 & 734  & 17.7 & 2.40 & 14.9 & 1.00 & 21.9 & 2.1 \\
2005/2006 & 1225 & 19.2 & 2.35 & 15.6 & 0.96 & 23.6 & 2.1 \\
\enddata
\tablenotetext{a}{The data were analyzed with the one-dimensional
analysis with $\alpha \leq 15^\circ$.}
\tablenotetext{b}{These results have been obtained by applying an
aperture cut of 0.22$^\circ$.}
\tablenotetext{c}{Results of the analysis of the
Gaussian ($\sigma =7.6'$)
smoothed ON and OFF sky maps.}

\end{deluxetable}

\begin{deluxetable}{clcccccc}
\tablecaption{Summary of the analysis
results of the {\TEV} data taken with the
Whipple Observatory 10~m telescope.
\label{tev2032}}
\tablewidth{0cm}
\tablehead{
\colhead{$\alpha_{2000}$ / $\delta_{2000}$} &
\colhead{S ($\sigma$)} &
\colhead{{\it ON}} &
\colhead{{\it OFF}} &
\colhead{{\it ON - OFF}} &
\colhead{$R_\gamma~ \rm (min^{-1})$} &
\colhead{Flux (Crab)}
}
\startdata
$20^h 32^m 27^s \pm 21^s_{stat} \pm 32^s_{syst}$,  & {{6.1}} & 
 {{9475}}
& {{8652}} & 
{{823}} &
{{0.19}} & {{0.08}} \\
$41^\circ 39' 17'' \pm 4'_{stat} \pm 6'_{syst}$ \\
\enddata
\end{deluxetable}

\clearpage



\begin{figure}
\epsscale{0.7}
\plotone{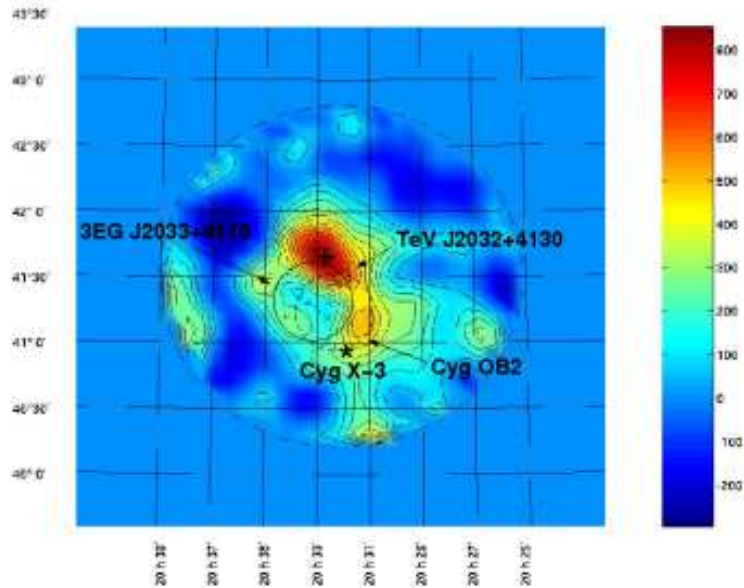}
\caption{
Gaussian-smoothed declination vs right ascension map of excess counts of 
the {\TEV} region (as in Figure~\ref{fg3}) overlaid with the position and extension
(where appropriate) of a 
number of astrophysical objects within the field of view as catalogued at 
other 
wavelengths. 
The small circle in the center of the field
of view represents the {\gr} source reported by HEGRA \citep{aha05a}. 
The approximate angular size of the high-density core of the Cygnus~OB2 
association (see \citealt{kns00}) is indicated by the dashed circle near the 
center of the field of view. The 95\% error circle of the EGRET source
3EG~J2033+4118 is also shown \citep{h99}. The cross marks 
the location of maximum signal
of the emission detected in the present work.
\label{fg4}}
\end{figure}

\end{document}